\documentclass[11pt,a4paper]{article}
\usepackage{amsfonts}
\usepackage{graphicx}
\usepackage{amsmath}
\usepackage[dvipdfm]{hyperref}
\usepackage{enumerate}
\RequirePackage{mathrsfs} \RequirePackage[sc]{mathpazo}
\RequirePackage{wasysym} \RequirePackage{setspace}

\makeatletter \@addtoreset{equation}{section}

\newcommand{\be}{\begin{equation}}
\newcommand{\ee}{\end{equation}}
\newcommand{\bea}{\begin{eqnarray}}
\newcommand{\eea}{\end{eqnarray}}
\begin{document}

\title{On  Mass Gap in Type IIB  Quantum Hall Solitons }
\author{  A. Belhaj$^{1,2,3,4}$,   M. J. I. Khan$^{1}$,  E.
H. Saidi$^{1,2,4}$, A. Segui$^{3}$\hspace*{-15pt} \\
\\
{\small $^{1}$ Lab. Phys. Hautes Energie:  Mod\'elisation et Simulation, Facult\'e des Sciences de  Rabat } \\
{\small  Universit\'e Mohamed V-Agdal, Rabat, Morocco }\\
{\small $^{2}$ Centre of Physics and Mathematics, CPM-CNESTEN, Rabat, Morocco  }  \\
{\small $^{3}$ Departamento de  F\'isica Te\'orica, Universidad de Zaragoza, E-50009-Zaragoza, Spain}\\
{\small $^{4}$Groupement National de Physique des Hautes Energies,
Si\`{e}ge focal: FSR, Rabat, Morocco }} \maketitle

\begin{abstract} We discuss the mass gap in quantum Hall solitons (QHS) embedded in superstring
theory. In particular, we give two holographic  models   which are obtained from   D-brane
 configurations in type IIB superstring compactifications.  The first one deals with the
 monolayered  system in the D3/D7  brane set up. The second  model   corresponds to 
a   multilayered  system which is described by   intersecting D5-branes wrapping  a  particular set of 3-cycles. In both models,  we have  shown that the mass  gap is related to the filling  factor.
\end{abstract}

\thispagestyle{empty} \newpage \setcounter{page}{1} \newpage

\section{Introduction}

\label{S:Introduction}
 Quantum Hall effect is one of the most interesting strongly  correlated electron system
 in condensed matter physics \cite{Wen}. At low energies, this  effect   which is
described by a topological field  theory  has been subject to great 
interest not only because of its experimental results but also from
its connection with  recent developments in string theory
\cite{BBST,ES,FLRT,BJLL}.\\ The quantum Hall states are characterized by
the filling factor $\nu$ describing  the ratio  between the
electronic density and the magnetic flux density.  When $\nu$ is a fractional
value,   it is called fractional quantum Hall effect (FQHE) describing interacting
electron systems. The first proposed series of the fractional
quantum states was given by
Laughlin and they are characterized by the filling factor $\nu _{L}=\frac{1}{%
k}$ where $k$ is an even integer for bosons and an odd
integer for fermions \cite{L}.  According to  Susskind \cite{S},  the  non-commutative  three dimensional   Chern-Simons
 gauge theory can 
 provide  a nice description  for  such Laughlin states. This framework   has opened a modeling  of these sort of phenomena in terms of
D-brane solitons of type II superstring theory.  In particular, it has been   shown that  the Susskind   model   can be constructed  by a  ten dimensional  
 system of  D0, D2
and D6-branes and F1 strings stretched between the D2 and D6-branes\cite{BBST}.  Using D-brane configurations on the K3 surface, a six dimensional 
type IIA stringy
realization of these type of models, has been  given in terms of D2 and D6-branes wrapping the K3
surface \cite{BS}.

On the other hand,  quantum Hall soliton (QHS) have  been 
 studied using  Anti de Sitter/conformal field theory (AdS/CFT) correspondence \cite{FLRT,BJLL}. In particular, 
 based on the Aharony-Bergman-Jafferis-Maldacena (ABJM) theory in   (2+1)-dimensions \cite{ABJM},   Chern-Simons (CS) 
 descriptions of QHS  have  been discussed in \cite{F,HLT}.
 These models are  embedded in the  3-dimensional  supersymmetric $N = 6$ CS quiver theory  with $\mbox{U(N)}_k\times \mbox{U(N)}_{-k}$ gauge symmetry. Recall that the ABJM proposal is
dual to M-theory propagating on $AdS_4 \times S^7/Z_k$, with an appropriate amount of fluxes, or equivalently to type
IIA superstring on $AdS_4 \times CP^3$.  Indeed, it has been shown that the FQHE
can be obtained from the world-volume action of the M5-brane filling $AdS_3$ inside $AdS_4$ space \cite{F}.
The corresponding  model has been derived from $d = 3$ flavored ABJM theory with the CS levels (1,−1).
Alternatively, a FQHE system in $AdS_4$/CFT$_3$ has been realized by adding fractional D4-branes, wrapping $CP^1$, to the ABJM theory \cite{HLT}. Moreover,  an extended model based on D6-branes wrapping  del Pezzo surfaces has been discussed in \cite{B,Be}.

A more generic  class of
FQHE models   which can   appear in real materials,  including graphene
\cite{Nobel,DSB}  is    characterized by  a vector $q_i$  and a real
symmetric invertible matrix $K_{ij}$ which are related  by the
filling factor \cite{Wen}. They   have a nice interpretation using solitonic
D-branes wrapping non trivial cycles in Calabi-Yau manifolds \cite{BS}. In
particular, the matrix $K_{ij }$ and the vector charge $q_i$ play an 
important role in the  quiver approach   of FQHS,  embedded in type II superstrings and M-theory,
compactified on deformed singular geometries that are classified by Dynkin diagrams of 
Lie algebras \cite{BS,BEFKSS,BFSS}.

The aim of this   work   is  to contribute to these activities by studding  the mass gap 
in QHS models, using type IIB D-brane configurations. First,  we give a
simple  model involving a 
 monolayer  system  based on  the D3/D7  brane set up. Then we  discuss  the case of the
 multilayered systems with  several   abelian gauge factors,
 using  the quiver method that is based on intersecting  D5-branes wrapping 3-cycles of the internal space. These 3-cycles  are given by a  line segment fibered by
intersecting spheres according to   extended Dynkin diagrams.   In both examples,
  the mass  gap   can be  related to the filling  factor of the corresponding QHS model. The last section is devoted to discussions.

The organization of the paper is as follows. In section 2 we review briefly the three dimensional gauge theory obtained from the  D3/D7 set up.
In section 3, we give a  stringy realization of the QHE in 1+2 dimensions from  such a  brane configuration, and then we discuss  the corresponding mass gap. In section 4, we study the mass gap in a hierarchical stringy description. This is obtained from quiver
gauge theories living on the world-volume of D5-branes that wrap a particular set of 3-cycles.


\section{The D3/D7 brane system}

\label{S:Quiver} The staring point is  the  (D3, D7) brane system that is embedded  in type IIB superstring\cite{Rey}. We consider
  D3-branes which  are extended in   (0123) directions. This leads to
four dimensional gauge interactions living on its world-volume. As we will see, the
three-dimensional fermions can be modeled by a stack of $k$
D7-branes oriented along (01245678) directions. The  system is
represented in the table \ref{t:one}.

\begin{table}[th]
\begin{center}
\begin{tabular}{|c|c|c|c|c|c|c|c|c|c|c|}
\hline
&0&1&2&3&4&5&6&7&8&9 \\ \hline
D3-brane & \checkmark & \checkmark & \checkmark & \checkmark &  &  &  &  &  &   \\ \hline
D7-brane & \checkmark & \checkmark & \checkmark &  & \checkmark & \checkmark &
\checkmark & \checkmark & \checkmark &   \\ \hline
\end{tabular}%
\caption{The D3/D7 brane configuration of the studied model}
\label{t:one}
\end{center}
\end{table}
Open strings have two-end points
satisfying  either the Neumann $(N)$ or Dirichlet $(D)$ boundary
conditions. One has three possibilities:  $(NN)$, $(DD)$ and $(ND)$.
These are 
completely fixed once the directions in which the D-branes live are determined.  In the above configuration, the D3-D7 open strings have
six $ND$ boundary  conditions. We know that the NS sector is
massive and only the R sector contains massless states. In this
case,  each D3-D7 pair gives a complex massless two-component spinor living in the  $2+1$ common directions.  This
configuration is non-supersymmetric and unstable because the branes
are repelled one from another in the direction $x_9$. The
excitations of RR open strings stretched between
D3 and D7-branes   give  massless fermions.   

The dual bulk description is obtained by taking a large number of
D3-branes and a finite number of D7-branes ( i.e., $N_f<<N_c$). In
this way,  the D7-branes can be treated as probes in the
near-horizon geometry, $AdS_5 \times S^5$, of the D3-branes. The D3-branes describe the four-dimensional gauge field dynamics, and
the D7-brane embedded in that background captures the
three-dimensional physics of the fermions. The corresponding
effective theory, which  has been used in \cite{Rey} to engineer
graphene multilayered, reads as
\begin{equation}
\label{rey}
S_4=\int\limits_{\mathbb{R}^{1,3}}\left[ \frac{1}{4g^{2}}F_{MN}^{2}+
\frac{1}{2g^{2}\xi }\left( \partial ^{M}%
A_{M}\right) ^{2}+L_f \right],
\end{equation}
where   $ A_{M }$ is the gauge field living on the D3-brane ($M=0,1,2,3$).
The second  term  on this equation is a covariant gauge fixing term, while the  last  one reflects the behavior of the fermions  localized  in the shared three
dimensions of the D3/D7 system. This term is  given  by  the following Lagrangian
\begin{equation}
L_f= \sum\limits_{a}\left( \overline{%
\psi }^{a}\gamma^{M }\left( i\partial_M-A_M\right) \psi
^{a}\delta \left( x^{3}\right) \right),
\end{equation}
where the $\gamma^{M}$ gamma matrices are given in terms of the
$2\times 2$ Pauli matrices. $\psi^{a}$ are the complex two-dimensional spinors and $a=1,\ldots,k$.  

The physics at the intersection of the
D7-branes and  D3-branes   can be described by an effective $2+1$-dimensional
field theory, with SO(3,2)  conformal invariance which  is a subgroup
of the SO(4,2)  symmetry of the D3 subsector. Following \cite{Rey}, the action (\ref{rey})
can be integrated out and reduced to
\begin{equation}
\label{rey2}
S_3=\int\limits_{\mathbb{R}^{1,2}}[ \frac{1}{4g^{2}}F_{\mu\nu}\frac{1}{\sqrt{-\partial^2}}F_{\mu\nu}+
\frac{1}{1+\xi }\left( \partial ^{\mu}%
A_{\mu}\right)\frac{1}{\sqrt{-\partial^2}}\left( \partial ^{\mu}%
A_{\mu}\right) +\left( \overline{%
\psi }^{a}\gamma ^{\mu }\left( i\partial_\mu-A_\mu\right) \psi
^{a}\right) ],
\end{equation}
where now $\mu=0,1,2$.

In the following sections, we will discuss the generation of a mass gap of the gauge field of the QHS  living on the three-dimensional intersecting IIB D-branes. In particular,  we will propose a relationship between the Kaluza-Klein mass gap and the filling factor $\nu$.  Then we extend this result by constructing a holographic model based on D5-branes wrapping non trivial 3-cycles.

\section{ Mass gap for D3/D7 brane quantum Hall solitons}
In this section, we study the mass gap for QHS  constructed in the D3/D7 brane set up. First we recall that  the simplest series of the fractional quantum states with the filling factor
$\nu=\frac{1}{k}$ can be described by a 3-dimensional U(1)
Chern-Simons theory coupled to an external electromagnetic field
$\tilde A$. The effective action of the system is
\begin{equation}
S_{CS}= -\frac{k}{4\pi }\int_{\mathbb{R}^{1,2}} A\wedge
dA+\frac{q}{2\pi }\int_{\mathbb{R}^{1,2}} {\tilde{A}}\wedge dA,
\label{sc}
\end{equation}
where $A$ is the dynamical gauge field and $q$ is the charge  of the
electron\cite{Wen,WZ}. 

Susskind  has conjectured that a two-dimensional quantum Hall fluid
of charged particles with the filling factor  $\nu=\frac{1}{k}$ can be
modeled by a non-commutative Chern-Simons  gauge theory at level
$k$ \cite{S}. This conjecture has opened a  new way  to apply
string theory  for  studying  low-energy systems in condensed matter
physics.
This has been based on the recent developments in string dualities, mainly AdS/CFT.  The first connection with string
theory was given by Bernevig, Brodie, Susskind and Toumbas
reproducing the QHE on the 2-sphere, $S^2$ \cite{BBST}. The
corresponding solitonic D-brane configuration involves a spherical D2-brane and
dissolved D0-branes on it. The system is placed in a
background of coincident D6-branes extended in the directions
perpendicular to the world-volume of the D2-brane on which the QHE
resides.  After a compactification on the K3 surface, this ten dimensional type IIA superstring construction can be reduced to 
a six dimensional one \cite{BS}. Other connections between string theory and QHS in $1+2$-dimensions were
found recently using the AdS/CFT correspondence \cite{FLRT,BJLL}.

The specific model that we deal with here is based on the D3/D7 brane system given in the previous section. In particular, we will see that the action (\ref{sc}) can be derived from the D3/D7 brane system in type IIB superstring. In type II superstring theory, the  Chern-Simons terms can be obtained from the Wess-Zumino (WZ) part of the action of a D-brane wrapping non trivial cycles and interacting with the R-R flux. Here, it will be shown  that it is possible to get the first term of (\ref{sc}) from the WZ action of the D3-brane. Similarly, the second part of (\ref{sc}) can be obtained from the interaction between the D3-branes and the D7-branes in the three dimensional shared space. This indicates how the D3-brane couples to the various backgrounds appearing in   type IIB superstring. Indeed, on the 4-dimensional world-volume of each D3-brane
one has a U(1) gauge symmetry. The corresponding WZ action reads as
\begin{equation}
\label{D31}
 S_{WZ} =\int\limits_{\mathbb{R}^{1,3}} C_{0}\wedge F\wedge F+C_{2}\wedge F+C_{4},
\end{equation}
where $C_{n}$  are  $n$-forms belonging to the R-R sector. 

To get the first part of the action (\ref{sc}), we consider a particular background given by the  vanishing condition $C_{2}= C_{4}=0$ on the D3-brane world-volume. In this case, the action (\ref{D31}) reduces to
\begin{equation}
 S_{WZ} = \int\limits_{\mathbb{R}^{1,3}} C_{0} \wedge F  \wedge F,
\end{equation}
where $C_{0}$ is the axion scalar field. After a simple integration by part, we get
\begin{equation}
 S_{WZ} =  - \int\limits_{\mathbb{R}^{1,3}} dC_{0} \wedge  A  \wedge F.
\end{equation}
Now we take a particular brane configuration, in which the $x^3$-direction is compact and the axion field $C_{0}$ behaves like $\frac{k}{2L}x^3$. $L$ is the size of the compact dimension $x^3\sim x^3+L$. This brane configuration can be supported by a stack of $k$ D7-branes
coupled to the gauge field living on the D3-brane. Integrating  over the $x^3$-direction, we obtain the  first Chern-Simons term of (\ref{sc})
\begin{equation}
\label{CS}
  S_{WZ} =-\frac{k}{4\pi } \int\limits_{\mathbb{R}^{1,2}} A\wedge dA,
\end{equation}  
where $\frac{k}{4\pi } = \int\limits_ {x_3} dC_{0}$. 

To obtain the full action of quantum Hall effect (\ref{sc}), one needs to couple the above D3-brane system to an external magnetic source. A new term containing the coupling between the external gauge field $\tilde{A}_{\mu }$ and the gauge potentials ${A}_{\mu }$ should be added.
Then, the last term of the effective Lagrangian  describing the $2+1$-dimensional interaction between the D3-branes and D7-branes that appears in (2.3) becomes
\begin{equation}
 \bar{\psi}^a\left[ \gamma ^{\mu } \left( i\partial _{\mu }%
 - A_{\mu }- {\tilde {A}}_{\mu }\right)  \right] \psi^a. 
\end{equation} 

Like the dynamical gauge field, that was obtained from the D3-brane world-volume, the external gauge field ${\tilde {A}_{\mu}}$  can  be  derived  using several brane configurations interacting with the R-R fields. However, for the external field we make use of the gauge field living on the D7-brane worldvolume. This field will play the role of  the  external source used in the quantum Hall description. On a D7-brane lives an  eight dimensional U(1) gauge field obtained from the quantization of the open string with the two ends on it. Wrapping this  D7-brane on $S^5$ we generate a three dimensional gauge field ${\tilde {A}_{\mu}}$. In  this way, this gauge field  ${\tilde {A}_{\mu
}}$ can be obtained 
 by reducing dimensionally the  gauge field  living in  eight dimensions down to three dimensions. This produces a direct coupling with the D3-brane gauge field because now the two fields live in the same space-time. The three dimensional coupling is given by the following action
\begin{equation}
 S\sim \int\limits_{\mathbb{R}^{1,2}}   \tilde{A} \wedge F.
\end{equation}
For $k$ D7-branes, the system is coupled to a U($k$) gauge fields $\tilde{A}$.  
Taking the  CS actions (3.7) and (3.5), integrating out the gauge field $\tilde{A}$ and using the equations of motion, one gets the following filling factor
\begin{equation}
\label{nu}
\nu=\frac{1}{k}.
\end{equation}

To obtain the relationship between the mass gap and the filling factor, we make use of the following steps. First, we add the Chern-Simons action (3.1) to the three dimensional  interacting action given in  (2.3)
\begin{equation}
 S_3+S_{CS}.
\end{equation}
Second, we take  the $\xi$ large limit. Then, by computing the powers of the derivatives of the dynamical gauge field $A$, we  find the form of the inverse of the propagator for the gauge boson in momentum space
\begin{equation}
kp-\frac{1}{g^2}p^2-p.
\end{equation}

Ignoring the Kaluza-Klein contribution, the spectrum has a mass gap given by
\begin{equation}
m\sim (k-1)g^2.
\end{equation}
Besides the three-dimensional  gauge coupling constant, the mass gap depends also on the order parameter $k$ responsible for classification of the various quantum Hall states. Using the relation $\nu=\frac{1}{k}$, the  mass gap takes the form
\begin{equation}
m\sim  \frac{1-\nu}{\nu}g^2.
\end{equation}
We see that the massless case is given by the line $\nu=1$ in the $(\nu, g)$ parameter space. This 
happens if the condition $k=1$ is satisfied, that is, the number of the D7-branes should  be one.

\section{Multi-layered systems} 
So far, we  have discussed the mas gap in a  QHS model with a single gauge group. In this section, we give a holographic model based on several gauge fields. In such a multi-layered system, we can
associate a separate gauge potential $A_\mu$ with each layer. This
is due to the fact that there is not tunneling and the current in each
layer is separately conserved. 

The most  general fractional quantum Hall system is described by the following Chern-Simons action
\begin{equation}
\begin{tabular}{ll}
$S\sim\frac{1}{4\pi }\int\limits_{\mathbb{R}^{1,2}} \sum_{i,j}K_{ij}A^{i}\wedge dA^{j}+2\sum_{i}q_{i}%
\tilde{A}\wedge dA^{i}$, &
\end{tabular}
\label{hd}
\end{equation}%
where $K_{ij}$ is a real, symmetric and invertible matrix ($\det K\neq 0
$) and where $q_{i}$ is a vector of integer charges. The external gauge field $\tilde{A}$
couples now to each current $\star dA^i$ with charge strengths $eq_i$, $e$ being the fundamental charge. The $K_{ij}$\ matrix and the $q_{i}$  charge vector in this effective field
action are very suggestive in  many physical senses. In the Wen-Zee model \cite{WZ}, $K_{ij}$ and $q_{i}$ are interpreted as order parameters of the model, classifying the various QHS states.   These parameters can be also explored to give a quiver gauge theory description of QHS
embedded in type II superstrings and  M-theory compactifications \cite{BS,BEFKSS,BFSS}.

Using eqs(\ref{hd}) and integrating over the gauge fields $A_{\mu
}^{i}$\ in the same way as in the Susskind model \cite{S}, we obtain the following filling
factor
\begin{equation}
\nu =q_{i}K_{ij}^{-1}q_{j},  \label{factor}
\end{equation}
generalizing eqs(\ref{nu}). This quadratic form can be split as
\begin{equation}
\nu=\sum_{i}K_{ii}^{-1}q_i^2+2\sum_{i<j}K _{ij}^{-1}q_iq_j.
\label{factora}
\end{equation}
From this equation we  can observe that different values of $\nu$  correspond to different forms for the matrix $K_{ij}$ and the
vector charge $q_i$. However, it is possible to have different models ($K_{ij}$ and $q_{i}$) with the same filling factor. 

From the string theoretic point of view, the matrix $K_{ij}$ can be
related with the intersection matrices of  the compactified manifolds,
considered as orbifold geometries \cite{FLRT}. The resolution of the
singularities of such geometries is nicely interpreted in terms of the Dynking
diagrams. In particular, the intersection matrix used for the
singularity resolutions is, up to some details, the opposite of the
Cartan matrices of the corresponding Lie algebras.  

In this section, we consider models which can be obtained from the IIB superstring living on
orbifolds like $AdS_5 \times S^5/\Gamma$, where $\Gamma$ is a discrete subgroup of the holonomy group of the dual geometry. The corresponding four dimensional gauge theory version, can be obtained by considering D5-branes wrapping the blown up two-cycles of ALE spaces
\cite{quiver1,LNV,quiver2,quiver3}. In this way, $N$ D5-branes are partitioned into $\ell$ subsets of $N_i$ D5-branes ($N=\sum_{i=1}^{\ell} N_i$), where $\ell$ is related to the
rank of the corresponding Lie algebras. The four-dimensional
gauge symmetry we will use in this section can be obtained by breaking the original symmetry to its abelian part involving only U(1) factor that describes the Coulomb phase of the theory. Using the quiver gauge theory results of \cite{BS}, the action (\ref{hd}) can be obtained  by wrapping D5-branes on particular 3-cycles.  We will see that these cycles can be viewed as a collection of 2-cycles fibered over a one dimensional compact base. 

To see how to get (\ref{hd}) using D5-branes, first we examine the model that gives a U(1) gauge theory. This will be subsequently extended to more general cases, with many gauge fields. A single D5-brane supports a U(1) gauge field. The corresponding DBI action gives (\ref{rey}) and must be supplemented with the WZ term that leads  to the Chern-Simom action. The first term of (3.1) can be obtained from the WZ action on a single D5-brane wrapping a particular 3-cycle. This three-cycle is embedded in a Calabi-Yau threefolds given by ALE spaces fibered over the complex plane. To be explicit we consider the ALE space with $A_1$ geometry which is identified with the cotangent bundle over the two dimensional sphere $S^2$. This complex geometry is 
\be
z_1^2+z_2^2+z_3^2= \mu,
\label{mu}
\ee 
where $z_i$ are complex variables and where $\mu$ is a complex parameter. The real part of this parameter is the radius squared of the sphere $S^2$, while the imaginary part can be identified 
with the $B$-field in superstring compactifications.  

The 3-cycle that we consider here will be obtained from a $A_1$ threefold, by varying the parameter $\mu$ over the complex plane parametrized by $w$. Consequently,  the equation (\ref{mu}) is given by 
\be
z_1^2+z_2^2+z_3^2= \mu(w).
\ee 
The 3-cycle can be given explicitly by a finite line segment with a $S^2$ fibration, where the radius $r$ of $S^2$ vanishes at the two interval end points and only there. This ensures that no more singular points appear. One way to realize this geometry is to use the  parametrization $r \sim \sin x$, where $x$ is a real variable parameterizing the interval $[0,\pi]$ in the $w$-plane.  This construction may be extended to more general geometries where we have intersecting  $S^2$'s  according to extended Dynkin diagrams. 

Having specified the 3-cycle, we will discuss the corresponding CS quivers of the compactified
type IIB superstring. The analysis here will be based on the WZ action on a single D5-brane  wrapping the 3-cycle $S^2 \times I_{[0,\pi]}$ previously constructed. In fact, on the six-dimensional world-volume of each D5-brane we can  have a U(1) symmetry with the following WZ action 
\begin{equation}
S_{WZ}=\int\limits_{\mathbb{R}^{1,5}}  F\wedge F \wedge C_2,
\end{equation}
where $ C_2$ is the R-R 2-form. Integrating by part, this WZ  action  becomes 
\begin{equation}
S_{WZ}=-\int\limits_{\mathbb{R}^{1,5}}  A\wedge F \wedge dC_2.
\end{equation}
Now we integrate the $dC_2$ three-form over $S^2 \times I_{[0,\pi]}$ and we find exactly the first part of the action (3.1), where $\frac{k}{4\pi } = \int\limits_ {S^2\times I_{[0,\pi] }} dC_{2}$.
Similarly, the second part of (3.1) is obtained from the following WZ term
\begin{equation}
\label{d52}
S_{WZ}=-\int\limits_{\mathbb{R}^{1,5}}  F \wedge C_4,
\end{equation}
where $C_4$ is now the R-R 4-from which couples to the D3-brane. To derive the external source $\tilde{A}$, the $C_4$  gauge field 
must be decomposed as follows
\begin{equation}
C_4 \to \tilde{A} \wedge \Omega,
\end{equation}%
where $\Omega$ is a harmonic 3-form on the compact space. After integrating $\Omega$ over $S^2 \times I_{[0,\pi]}$, the equation  (\ref{d52}) takes the form
\begin{equation}
q \int\limits_{\mathbb{R}^{1,2}} \tilde{A}\wedge F.
\end{equation}%
Now, the gauge field $\tilde{A}$ is interpreted as a magnetic external source interacting with the D5-brane system system.  

We can follow the same analysis to get a general solution with an arbitrary number of 3-cycles. However, the general study is beyond the scope of the present work, and we will restrict ourselves to only two isolated  three-cycles ($K_{i<j}=0, (i,j=1,2)$). In this case, (\ref{factora}) reduces to
\begin{equation}
\nu=\sum_{i}K_{ii}^{-1}q_i^2.
\label{factorb}
\end{equation}
The explicit model we deal with here represents a two layers QHS, and it is associated with a
$\mbox{U(1)} \times\mbox{U(1)}$ quiver gauge theory with the following $K_{ij}$ matrix
\begin{equation}
\left(
\begin{array}{cc}
 k_1 & 0\cr
0&k_2%
\end{array}%
\right). \label{matrix}
\end{equation}%
This  matrix can be considered as an extended Cartan
matrix involving two simple roots. Geometrically, this may  correspond to two isolated
Riemann surfaces with a positive genus (instead of $S^2$'s). If we take the charges values $q_i=(1,1)$, then the filling factor of (4.11) is
\begin{equation}
\nu=\frac{1}{k_1}+ \frac{1}{k_2}.\label{factorb}
\end{equation}
As we have done in the previous section, and using the fact that $\nu=\sum_i\nu_i=\nu_1+\nu_2$, we get the following mass gap formula for the bi-layer system
\begin{equation}
m\sim \sum_i \frac{1-\nu_i}{\nu_i}g_i^2.
\end{equation}
In string theory compactification, the gauge coupling $g_i$ is related to the volume $V_i$ of
the cycles on which the D5-branes are wrapped, $g_i\sim \frac{1}{V_i}$. We see that the mass gap can also be related to the closed string moduli. We believe that this connection deserves to be studied further.
\section{Discussions}

In this  work, we have discussed  the mass gap for quantum Hall
solitons which have been constructed from string theory compactification. We have shown that the mass gap can be related to the filling factor of the quantum Hall systems. In particular, we have obtained two simple holographic models for quantum Hall
solitons that are based on D-brane configurations in type IIB superstring. The first one has been constructed from the monolayer system based on the D3/D7 brane configuration  by adding Cherm-Simons terms. The second model is associated with a quiver gauge theory describing bi-layered system constructed with intersecting D5-branes wrapping on a particular set of 3-cycles. This analysis can be generalized to multilayered systems by considering an arbitrary number of three-cycles with arbitrary charges. A simple generalization can be obtained from
geometries whose intersection forms may be represented by $K_{ij}=k_i\delta_{ij}$.

\emph{Acknowledgments:}  We thank  URAC 09/CNRST.
 We thank  also the   grant
A/031268/10. MJIK  cordially thanks for the financial support
provided by Prof. H. Saidi to engage him for teaching in university
as honorary lecturer.   He would like to thank his Parents who make
him enable to carry research with patience. AS has been supported by the grant FPA2009-09638.

\end{document}